\documentclass{article}
\usepackage[nonatbib, final]{neurips_2022}
\usepackage{cite}
\usepackage{amsmath,amssymb,amsfonts}
\usepackage{amsthm}
\usepackage{algorithm}
\usepackage{wrapfig}
\usepackage{algpseudocode}

\usepackage{graphicx}
\usepackage{textcomp}
\usepackage{xcolor}
\usepackage{subcaption}
\usepackage{multirow}
\usepackage{booktabs}
\usepackage[utf8]{inputenc} 
\usepackage[T1]{fontenc}    
\usepackage{hyperref}       
\usepackage{url}            
\usepackage{booktabs}       
\usepackage{amsfonts}       
\usepackage{nicefrac}       
\usepackage{microtype}      
\usepackage{xcolor}         

\title{Zero Day Threat Detection Using Graph and Flow Based Security Telemetry}

\author{Christopher Redino,
        Deloitte\thanks{Deloitte and Touche LLP}\And
        Dhruv Nandakumar,
        Deloitte${^*}$\And
        Robert Schiller,
        Deloitte${^*}$\And
        Kevin Choi,
        Deloitte${^*}$\And
        Abdul Rahman,
        Deloitte${^*}$\And
        Edward Bowen,
        Deloitte${^*}$\And
        Matthew Weeks,
        Deloitte${^*}$\And
        Aaron Shaha,
        Deloitte${^*}$\And
        Joe Nehila,
        Deloitte${^*}$\And
}

\begin{document}

\maketitle

\begin{abstract}
Zero Day Threats (ZDT) are novel methods used by malicious actors to attack and exploit information technology (IT) networks or infrastructure. In the past few years, the number of these threats has been increasing at an alarming rate and have been costing organizations millions of dollars to remediate. The increasing expansion of network attack surfaces and the exponentially growing number of assets on these networks necessitate the need for a robust AI-based Zero Day Threat detection model that can quickly analyze petabyte-scale data for potentially malicious and novel activity. In this paper, the authors introduce a deep learning based approach to Zero Day Threat detection that can generalize, scale, and effectively identify threats in near real-time. The methodology utilizes network flow telemetry augmented with asset-level graph features, which are passed through a dual-autoencoder structure for anomaly and novelty detection respectively. The models have been trained and tested on four large scale datasets that are representative of real-world organizational networks and they produce strong results with high precision and recall values. The models provide a novel methodology to detect complex threats with low false-positive rates that allow security operators to avoid alert fatigue while drastically reducing their mean time to response with near-real-time detection. Furthermore, the authors also provide a novel, labelled, cyber attack dataset generated from adversarial activity that can be used for validation or training of other models. With this paper, the authors' overarching goal is to provide a novel architecture and training methodology for cyber anomaly detectors that can generalize to multiple IT networks with minimal to no retraining while still maintaining strong performance.  
\end{abstract}

\section{Introduction}

Increased complexity in organizational attack surfaces and data volumes lead to a constant improvement in attack techniques and artificial intelligence (AI)-driven processes for cyber attacks \cite{threatLandscape}. Cyber Threat Detection Solutions play a pivotal role in  proactively identifying and responding to these threats \cite{IDSOverview} in a manner that can prevent costly damages to organizations \cite{IBMCost}, that could cost a business an average of over \$1 billion in long-term cost \cite{powers_fincher_silber}. Today's petabyte-scale security telemetry and attack sophistication quickly overcome limitations of traditional Signal-based Intrusion Detection Systems (SIDS) \cite{IDSOverview}, and require more and more robust AI-based detection mechanisms to identify hard-to-detect threats. In addition, these systems are largely limited to rules based alerting mechanisms constructed against indicators of compromise (IoC) gathered from threat research and intelligence. A well known limitation of rules based detections is their inability to match to new threats (i.e. zero day threats) that may not match the specific rule. This increasing complexity coupled with the inherent weakness in native rules based IDS engines allows attackers to continuously discover novel ways to exploit systems. Currently, ZDTs have an average growth in number of 125\% year-over-year since 2015 \cite{symantecReport}. 

Henceforth, ZDTs are novel Tactics or Techniques under the MITRE ATT\&CK Framework (https://attack.mitre.org) that have not been encountered in threat research / intelligence previously (i.e. no known rule presently exists to match for the threat). In this paper, we propose a novel dual-autoencoder approach to ZDT detection utilizing network flow telemetry augmented with asset-specific graph features to contextualize said flows. The overarching goal of the proposal is to build a robust anomaly detection mechanism that can identify ZDTs in near real-time while simultaneously being able to generalize across different network topologies with minimal retraining.  In this work we also place a heavy emphasis on producing models with a very low false positive rate and high recall in order to minimize Alert Fatigue on the security operator while delivering actionable results. The key contributions of this paper are:
\begin{itemize}
  \item Proposing and implementing a dual-autoencoder approach for network anomaly detection and attack novelty detection successively.
  \item Utilizing asset-level graph features in our deep learning architecture alongside flow-level features.
  \item Comparing the performance of traditional single-model methods without flow features with our method to evaluate the models' ability to generalize across networks.
  \item Providing a novel, labelled attack dataset consisting of network flow information for use in model development, benchmarking, and further research.
\end{itemize}

The paper begins with a literature review discussing previous work in the ZDT detection field so far, followed by a description of training methodology, implementation details, and results. We then conclude with remarks on some of our key design decisions, and description of future work.

\section{Review and Shortcomings in the Literature}

\subsection{Review of the Literature}
Our literature review will focus on three distinct components - proposed methods for ZDT detection, Autoencoders (AEs) for cyber anomaly detection, and the use of graph features and metrics in accomplishing the same. 

While cyber threat detection for known attacks such as Botnet infections \cite{botnetDetection} and other such attacks has been extensively studied in literature, ZDTs have only recently gained traction in terms of research. Zhao et al.\cite{transfer1} and Sameera et al.\cite{transfer2} utilized Transfer Learning to identify ZDTs, and Abri et al.\cite{transferGAN} and Kim et al.\cite{ZeroDL} also experimented with Deep Learning for ZDT detection utilizing methods such as Generative Adversarial Networks. There has also been other work in detecting ZDTs using methods such as, but not limited to, rule-based approaches as well as unsupervised clustering methods using malware opcode sequences \cite{ruleSIEM}, \cite{opcode} and Bayesian Classifiers for ZDT detection \cite{baysean}. However, there has also been a predominant focus on identifying ZDTs using network flow telemetry, similar to our paper. Lobato, Lopez, et al. \cite{adaptiveZero}  utilize supervised machine learning approaches such as Support Vector Machines (SVM) and Stochastic Gradient Descent (SGD) Algorithms to classify network flow telemetry as malicious or benign using an adaptive data modeling and pipeline approach with promising results, achieving precision and recall values between 65.7\% and 97.3\%. Blaise et al.\cite{portbased} utilized network flow telemetry to identify ZDTs using an unsupervised approach that identifies anomalous port usage. Sarhan et al.\cite{zeroShot} also utilized network flow telemetry to identify ZDTs using a Zero-Shot Learning approach with Random Forest (RF) and Multi-layer Perceptron (MLP) models using a novel data splitting approach to hold out attack classes as ZDTs to measure performance. While utilizing supervised or statistical approaches to ZDT detection has been shown to have some success, we believe that the lack of accurately labeled training datasets and the vast differences between network topologies of various organizations could cause difficulties in either successful model training or future generalization. 

Autoencoders are powerful tools in anomaly detection, particularly in the cyber domain, due to the fact that they can be trained solely on benign data and identify anomalous examples using a threshold on an erroneous reconstruction loss. Hindy et al.\cite{vanillaAutoEnc}, Yousefi-azar et al.\cite{yousefiAE}, and An and Cho\cite{VAE}  demonstrate the effectiveness of Network Intrusion Detection using Autoencoders trained on network flow telemetry in papers,and achieve strong precision and recall values compared to traditional methods such as RF models. Zhang et al.\cite{semanticAE} also utilize a modified Autoencoder approach with semantic segmentation of attack types in natural language in to identify ZDTs with strong accuracy values up to 88.3\% as compared to other methods. However, the above studies are all limited to single Autoencoder models performing anomaly detection after being trained on benign data, and only utilize flow-based features or semantic representations of said features.

We believe that graph-based features could help augment network flow telemetry and allow anomaly and ZDT detectors to generalize better across networks. By extracting expressive representations such that graph anomalies and normal objects can be easily separated, or the deviating patterns of anomalies can be learned directly through deep learning techniques, graph anomaly detection with deep learning (GADL) is starting to take the lead in the forefront of anomaly detection \cite{graphSurvey}. Yu, Cheng, et. al\cite{yu2018netwalk} had success utilizing deep learning to produce dynamic network embeddings for anomaly detection. However, their approach required utilizing embeddings for entire network topoligies over time to detect anomalies rather than our proposed approach for utilizing graph-based features only for assets involved in a network flow. Graph-based methods that do not utilize deep learning have also been seen to yield strong results, e.g. Chen et al.\cite{spectralDecomp} utilize Multi-Centrality Graph Spectral Decomposition on Intrusion Detection (ID) data from the University of New Brunswick along with k-means clustering to demonstrate that graph features can be utilized to identify network intrusions. Furthermore, Sadreazami et al.\cite{sadreazamiGraph}, Eric Dull\cite{dullGraph},  and  Noble and Cook\cite{nobleGraph} also utilized graph based statistical techniques to identify cyber anomalies in network flow information. We wanted to extend the applicability of graph-based cyber anomaly detection by using asset-level graph features in AEs along with network flow features to enhance anomaly detection capabilities and improve our models' ability to generalize.
 
\subsection{Shortcomings in Testing Methodology and Data in Literature }

The majority of studies conducted previously rely on one or two primary datasets of CICIDS2017, NSL-KDD, or KDD-99 for training and evaluation of models \cite{adaptiveZero} \cite{vanillaAutoEnc}, \cite{yousefiAE}, \cite{semanticAE}. Some studies also utilize other open-source datasets such as the MAWI Wide Archive \cite{portbased} or datasets such as the UNSW-NB15 \cite{zeroShot}, and have aimed to create sophisticated machine learning methods for cyber anomaly detection on a per-network basis. However, we believe that adopting the approach of having to retrain our ZDT detector during deployment on new industry architecture will prove to be infeasible due to the scale of the networks and the volume of security telemetry. Furthermore, retraining our models on production data could also involve training on anomalous network flows from Advanced Persistent Threats (APTs) that could be learned by our models, and consequently not identified as ZDTs. Our hypothesis is that in order to improve the ability of our models to generalize across networks, we would need to encode information about not only the network flows themselves, but also about the assets in the network involved in those flows. As a consequence, our team did not utilize commonly available datasets such as NSL-KDD or CICDS2017 due to the absence of asset identification information (source and destination IP addresses). Instead, we utilized larger datasets with asset identifiers such as the National Collegiate Cyber Defence Competition (NCCDC) 2020 and 2021 dataset and the MAWI Archive as detailed below. 

\section{Methodology}
In this section, we will cover the key concepts related to the datasets we utilized for training and evaluation, our high-level approach to ZDT detection, and our feature engineering. 

\subsection{Datasets}
Our approach to dataset selection was based on three criteria. Namely, the datasets must include flow source and destination information such as IP addresses, they must include basic flow-level features such as flow duration, destination port, timestamp, etc., and we must be able to collect a variety of datasets that vary in size and time-frames. Consequently, we relied on four primary datasets for training and evaluation, listed in \ref{table:1}. As seen in \ref{table:1}, we are also introducing a novel dataset (NCCDC) for the purposes of model development, benchmarking, and further research. NCCDC was generated using packet captures from the United States National Collegiate Cyber Defence Competition \cite{williamsNCCDC}. The dataset was captured during red-team and blue-team activities where the red-teams conducted one of four attacks: Scanning, Interrogation, Botnet, or Command and Control. The dataset we provide consists of network flow features labelled as one of the four attack types mentioned previously or a benign flow. 

\begin{table*} [t!]
    \centering
    \caption{Datasets Used For Training and Evaluation}
    \label{table:1}
\begin{tabular}{|p{0.1\linewidth}|p{0.35\linewidth}|p{0.45\linewidth}|}
    \hline
    Dataset & Description & Notes \\
    \hline \hline
    MAWI Archive (MAWI) & Real network PCAP data from hundreds of devices across 12 universities in Japan. & Flow data for 7 consecutive days from 2021, and 1 day from 2016 were used from a dataset of over 14 years. Dataset is not labelled but is considered benign due to the extreme class imbalance of anomalies in real-world networks.  \\
    \hline
    Ntnl. Collegiate Cyber Defence (NCCDC) & Consists of labelled blue-team and red-team data from real attack simulations on a cyber range. & Data spanned two consecutive days in 2020 and in 2021. Attack types consist of Network Scanning, Interrogation, Botnet, and Command and Control. \\
    \hline
    Deloitte Internal Flow (DIF) & Data consists of real Deloitte internal network traffic from 12 Deloitte US and US-India locations. & Flow data for 7 consecutive days was used. Dataset is not labelled but is considered benign due to the extreme class imbalance of anomalies in real-world networks. \\
    \hline
    Deloitte Codex Malware Lab (Codex) & Data consists of network data of over 30 million flows over 1 year from hundreds of real malware sample detonations in Deloitte's internal malware cyber-range. & Dataset is labelled with malware class name, and was correlated with threat intelligence to extract higher level attack-type labels including Botnets, Ransomware, Infostealer, etc. Data did not contain non-malicious flows.\\
    \hline \hline
\end{tabular}
\end{table*}

\subsection{Feature Engineering}
Our ZDT detector relies only on network flow data to identify potentially malicious activity on organizational networks. Our base dataset was specifically chosen to maximize compatibility with most security logging and monitoring (L\&M) infrastructures and requires only connection level information such as Connection Source and Destination IP addresses, connection duration and timestamp, port, protocol, and the total size of information exchanged between source and destination. We also utilize all of the network flow data to construct a network interaction graph of the network which we then use to derive graph-features for the various flows. The network graph is constructed by assigning a node to every unique IP address present in the dataset and an edge with weight 1 for every unique flow between two nodes. If there exist multiple flows, the edge weight is incremented by 1 for every flow present. Once the graph is constructed, we extract the graph features for the source and destination IP address involved in a flow, for every flow including  Degree Centrality and Pagerank \cite{xing2004weighted}, Clustering Coefficient \cite{clusteringCoeff} and Betweenness Centrality \cite{freemanBetween}, In-degree, Out-degree, In-weight, Out-weight, Hub and Authority values using the HITS algorithm \cite{HITSPaper}, and Cross-community flow detection using Label Propagation \cite{asyncLP}. Once the graph features are calculated, each flow is represented by 25 unique features (before preprocessing steps) per flow, which is inclusive of the graph features and flow level features. It is important to note that the source and destination IP addresses are not used as features to feed our model, and are only used to construct our network graph. 

\begin{figure*}[h]
    \includegraphics[width=\textwidth]{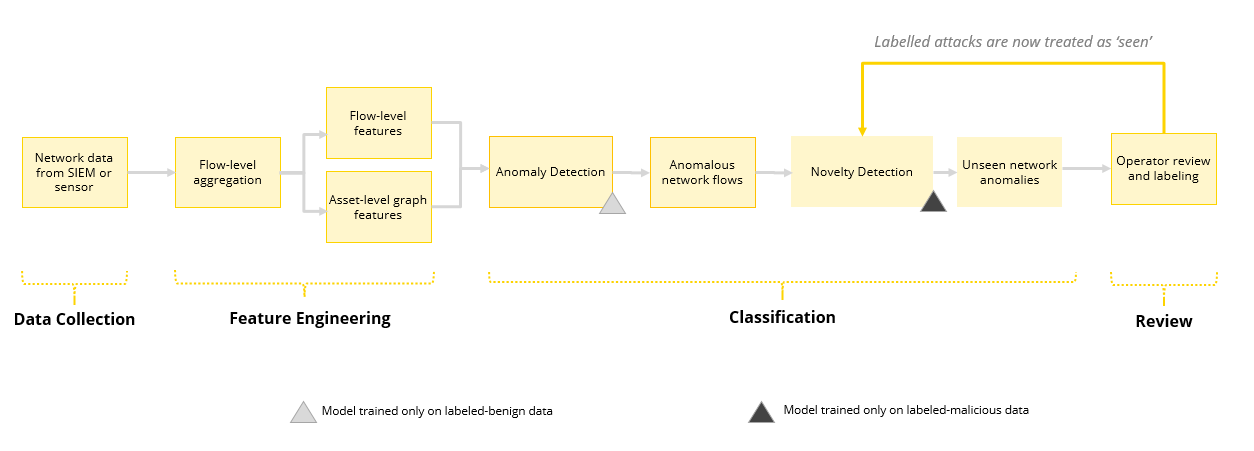}
    \centering
    \caption{High-Level Architecture}
    \label{fig:modelArch}
\end{figure*}

\subsection{ZDT Model Architecture}
Our ZDT detection architecture is a workflow that consists of two independently trained AEs that perform the following functions:
\begin{enumerate}
    \item \textbf{Anomaly Detector (AD)}: An AE trained solely on benign network traffic, responsible for identifying anomalous and potentially malicious flows, indicated by a high reconstruction loss. 
    \item \textbf{Novelty Detector}: An AE trained only on 'known' or previously seen attack types, responsible for identifying novel or zero-day attacks, indicated by a high reconstruction loss. 
\end{enumerate}

During inference, network flows are normalized and passed through the AD to identify if the network flow is suspicious. If the reconstruction losses are high, the original flows are then re-normalized using the normalization parameters for the Novelty Detector and passed through it to identify if the network anomaly is novel or not. Novel threats are then sent to security operators for review and slated to be used for retraining our models based on operator review. The workflow is also depicted in \ref{fig:modelArch}. 

\section{Experimental Design}
\subsection{Performance Metrics and Baseline Modeling}

Due to the highly imbalanced nature of real-world cybcersecurity datasets and their tendency to be heavily skewed towards benign events, we utilized \textbf{Precision, Recall} and the area under a Receiver Operating Characteristics (ROC) curve (\textbf{AUC}) to evaluate our modes. Furthermore, we also worked with the Deloitte Detect and Respond cyber unit and threat hunt teams to establish broad benchmarks we believe our models should accomplish to be useful in a practical setting. Our models should be able to make predictions in near-real-time with a high precision (above 0.9), a recall as high as possible (above 0.8) and maximize the AUC. Based on our PMs, our team, in collaboration with other data scientists at Deloitte's A.I. Center of Excellence created the following model implementation and evaluation framework. To establish baseline peformance, we first trained a Random Forest Multi-Class Classifier (RF) \cite{breiman2001random} and a a Multi-Class Support Vector Classifier \cite{Boser92atraining} to classify network flow data as benign, or as a distinct attack type. Although they cannot be used to detect novel threats in this manner, we wanted to be able to estimate problem difficulty, which showed low performance with both the RF and SVM struggling to classify labels with low support (0.7 and 0.17 AUC respectively). Next, we trained a single Dense AE on benign network traffic to observe whether it would be able to identify anomalous network traffic on the basis of poor reconstruction loss. This method has had wide success in literature and we wanted to ensure we could recreate similar results with our datasets. Similarly, we then trained a single Dense AE on benign network traffic as well as network traffic of three distinct attack types. We then computed reconstruction loss on a fourth independent attack type to observe if this fourth attack would be flagged as anomalous by the AE. The goal would be to test if the model would be able to identify ZDTs effectively, which in this case would be the distinct fourth attack type.

\subsection{Anomaly Detection}
The goal of the model is to separate benign and malicious traffic. We trained the AD on benign network traffic, with only 84 network flow features extracted from packet capture files using the CICFlowMeter tool \cite{CICFLowMeter}, from a single network and assess reconstruction loss on malicious traffic from that network, as well as benign traffic from two other networks. This would test if the model can generalize across multiple networks but still identify anomalous traffic. Next, we trained the AD on benign traffic from two networks, with only network flow features, and validate on malicious traffic from network, and benign traffic from the third to see if there is a boost in performance. We repeat this experiment while training with benign traffic with three networks as well.Finally we performed the above two sets of experiments with a Dense AE, but this time, we included not only network flow features but also asset-based graph features discussed in III B during training and validation.  Based on these experiments, we could then assess if graph features provide the necessary asset level context to allow our models to generalize between networks.

\subsection{Novelty Detection}
The goal of the Novelty Detector would be to analyze anomalous or malicious traffic and identify if the anomaly is of a class that has been previously seen or if it is 'novel', i.e. a zero-day attack. Consequently, the Novelty Detector  is trained solely on malicious network traffic and evaluated based on how it reconstructs examples of attack types similar to ones it has been trained on and 'holdout' attack types that are fundamentally different from any examples seen during training time. Holdout attack types are curated to be tactically and technically different from attack types shown to the model during training to ensure that they are indeed representative of ZDTs rather than different variations of a parent attack type. Furthermore, we also set aside examples of different attacks that are part of the same attack family to ones that are seen during training, e.g. different Botnet infections, to determine that our Novelty Detector does not incorrectly classify different kinds of procedures under the same tactic or technique as novel attacks. Experiments were also conducted to determine that novelty detection could be generalized across networks by determining that an attack type in a previously unseen network is not flagged as novel if the model has seen a similar example from a network during training. We also conducted experiments to determine the converse applies, where similar novel attack types are flagged as novel regardless of what network they originated in.

\subsection{End-to-End Pipeline Performance}
Once the AD and Novelty Detector were benchmarked and tested individually, we then passed a mixture of benign and anomalous examples through an end-to-end pipeline containing both models sequentially in order to measure performance. The data were first passed through the AD, and malicious events flagged by that model were then passed through the Novelty Detector to identify if they were ZDTs. Overall precision, recall, and AUC values were then computed. 

\section{Implementation and Results}

\subsection{Autoencoder Trained on Benign Data}
Training and validation of these models was also performed using data from the NCCDC and MAWI datasets. The NCCDC dataset was first split into benign and malicious data based on the label of the example. The benign data was then split into training (NCC-train) and testing (NCC-test) datasets with the training dataset containing 70\% of randomly sampled benign examples. NCC-train was normalized using Min-Max normalizaiton, and NCC-test was normalized using NCC-trains normalization parameters. The malicious data from the NCCDC dataset (NCC-holdout) was also normalized using NCC-train's normalization parameters.

\begin{figure}[h!]
\caption{Baseline Loss Histograms}
\centering
\begin{subfigure}{.5\textwidth}
    \centering
    \includegraphics[width=7cm, height=4cm]{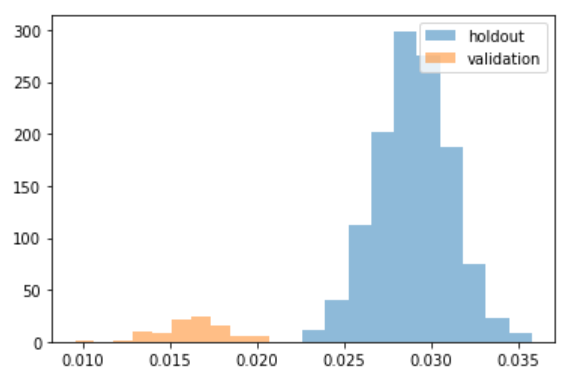}
    \caption{AE Reconstruction Loss - Trained on Benign}
    \label{fig:AEBaselineBen}
\end{subfigure}%
\begin{subfigure}{.5\textwidth}
    \centering
    \includegraphics[width=7cm, height=4cm]{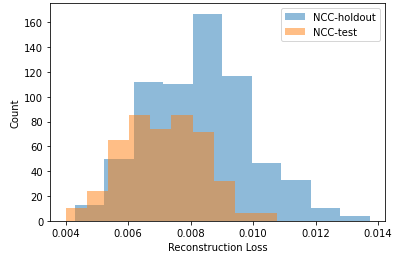}
    \caption{AE ZDT - Loss Histogram}
    \label{fig:AESingleCombinedHist}
\end{subfigure}
\end{figure}

The AE was then trained on NCC-train and performance was evaluated on NCC-test and NCC-holdout. A strong model will produce lower reconstruction loss for Benign examples than for Malicious examples. Our experiments with NCCDC supported this hypothesis, with our trained model able to distinguish between Benign and Malicious events without error (\ref{fig:AEBaselineBen}) at a reconstruction threshold of 0.022. This yielded a perfect AUC of \textit{1} which provided a significant boost in performance over the RF and SVM.

\subsection{AE Trained on Benign and Malicious Data}
Similar to the previous experiment, NCC-train and NCC-test sets were created using \textit{Benign, Scanning, Interrogation, and Command and Control} examples from NCCDC. \textit{Exfiltration} examples were used as the NCC-holdout dataset. Our hypothesis was that, if the AE is a strong ZDT detector, it would have produced higher reconstruction losses on NCC-holdout than on NCC-test given that NCC-holdout contained a novel, fundamentally different, attack type. 

However, results from the experiment indicated that a single AE architecture could not effectively distinguish between novel and previously seen attack types with an AUC of 0.72 and significant overlap in the loss histogram\ref{fig:AESingleCombinedHist}. These results supported our hypothesis that AEs, while strong anomaly detectors,  \textbf{would not be effective ZDT detectors} in practice, especially as the number of known attack types starts to rise. These experimentation results formed the basis for our decision to move to a two AE structure: the AD and the Novelty Detector.

\subsection{The AD}
Training of the AD was performed using only benign data from three datasets from distinct and independent networks: NCCDC, MAWI 2021, and MAWI 2015. MAWI 2015 and MAWI 2021 are assumed to be independent as both datasets have different characteristics in network flow values. In each of the three experiments outlined below, each dataset was normalized independently of other datasets using Min-Max normalization. Furthermore, malicious data from the NCCDC dataset were treated as holdouts during training and testing, and were normalized separately using normalization parameters acquired from normalizing NCCDC benign data. It is also worth noting that, unless otherwise mentioned, training and test splits for training data are randomly sampled with a 70\%/30\% split respectively.

\begin{figure}[h!]
\caption{AD Performance}
\centering
\begin{subfigure}{.4\textwidth}
    \centering
    \includegraphics[width=6cm, height=4cm]{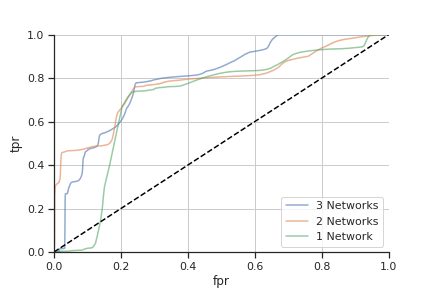}
    \centering
    \caption{AD ROC for Different Numbers of Training Networks}
    \label{fig:roc_nograph_anomaly}
\end{subfigure}%
\begin{subfigure}{.5\textwidth}
    \caption{AUC Scores for AD Trained on Different Numbers of Networks}
    \centering
\begin{center}
\begin{tabular}{|c|c|c|} 
  \hline
  No. of Networks & Flow Only & Flow and Graph \\ 
  \hline
  1 & 0.72 & 0.993 \\ 
  2 & 0.77 & 0.999\\
  3 & 0.79 & 0.999\\
  \hline
\end{tabular}
\end{center}
\label{fig:anomaly_results}
\end{subfigure}
\end{figure}

The first experiment tested the AD's ability to generalize to multiple networks using only network flow features, and consisted of three runs. In the first run, the model was trained on benign data from the NCCDC dataset and tested on NCCDC and the two MAWI datasets. Next, the training data from NCCDC was supplemented with benign traffic from MAWI 2015, and the model was evaluated against all three networks. Finally, the model was trained and evaluated using data from all three networks. The models trained on data from one, two, and three networks generated AUC scores of \textit{0.72}, \textit{0.77}, and \textit{0.79}, respectively, indicating stronger generalization with additional training data from distinct networks (\ref{fig:roc_nograph_anomaly}).

The second experiment followed a similar setup, but with the addition of asset-level graph features and only utilizing 6 foundational network flow features identified in III B instead of 84 features. Subsetting the network flow features while seeing if we can still maintain performance will allow our models to be more flexible with deployment without the need for specialized sensors on networks.  In the first run, the network was trained on NCCDC and evaluated on NCCDC, MAWI, and DIF. Next, the training data was supplemented with MAWI and the model was again evaluated on all three networks. Finally, the model was trained and evaluated on all three networks. In all three runs the models demonstrated AUC greater than \textit{0.99}, with small but consistent performance increases from training with additional networks. The ROC curves are not shown because they are nearly identical, but a summary of the results of the AD experiments can be found in \ref{fig:anomaly_results}. This highlights that our AE approach combined with asset-level graph features demonstrated near perfect performance in generalize to new networks as an anomaly detector; significantly higher than an AE with only network flow features.

\subsection{The Novelty Detector}

The novelty detector was trained on the NCCDC and Codex datasets using 6 flow-level features along with asset-level graph features. In each training run, one attack class was held out and treated as a ZDT at evaluation, and the loss threshold was selected to maximize precision while maintaining some sensitivity to positive examples (recall generally greater than $0.50$). The models were able to discriminate between novel and known attacks, though the performance varied significantly with the choice of holdout (\ref{fig:novelty_results}). In the test datasets, holdout attacks represented less than $ 2\% $ of all events, mimicking the rarity of ZDTs. This class imbalance accounts for the discrepancy between the measured AUC and precision-recall.

\begin{figure}[h!]
\caption{Novelty Detector and End-to-End Performance}
\centering
\begin{subfigure}{.5\textwidth}
    \caption{Performance of Novelty Detector}
    \centering
\begin{center}
\begin{tabular}{|c|c|c|c|} 
  \hline
  Attack & AUC & Precision & Recall \\ 
  \hline
  rat & 0.84 & 0.69 & 0.45 \\
  infostealer & 0.94 & 0.56 & 0.29 \\
  command/control & 0.91 & 0.97 & 0.83 \\
  ransomware & 0.95 & 0.80 & 0.48 \\
  botnet & 0.95 & 1.00 & 0.89 \\
  interrogation & 0.89 & 0.91 & 0.92 \\
  worm & 0.97 & 0.83 & 0.53 \\
  downloader & 0.86 & 0.71 & 0.54 \\
  scanning & 0.84 & 0.93 & 0.91 \\
  \hline
  \textbf{Average} & \textbf{0.91} & \textbf{0.82} & \textbf{0.65} \\
  \hline
\end{tabular}
\end{center}
\label{fig:novelty_results}
\end{subfigure}%
\begin{subfigure}{.5\textwidth}
    \centering
\caption{End to End Performance of Two-Model Approach}
\begin{center}
\begin{tabular}{|c|c|c|c|} 
  \hline
  Attack & AUC & Precision & Recall \\ 
  \hline
  rat & 0.84 & 0.71 & 0.45 \\
  infostealer & 0.94 & 0.60 & 0.28 \\
  command/control & 0.91 & 0.98 & 0.83 \\
  ransomware & 0.87 & 0.78 & 0.40 \\
  botnet & 0.96 & 1.0 & 0.89 \\
  interrogation & 0.90 & 0.92 & 0.92 \\
  worm & 0.97 & 0.83 & 0.50 \\
  downloader & 0.86 & 0.74 & 0.53 \\
  scanning & 0.84 & 0.93 & 0.89 \\
  \hline
  \textbf{Average} & \textbf{0.90} & \textbf{0.83} & \textbf{0.63} \\
  \hline
\end{tabular}
\end{center}
\label{fig:e2e_results}
\end{subfigure}
\end{figure}

\subsection{End to End Testing}

To evaluate the overall performance of the two-model approach, a separate test was conducted for each attack class, treating it as a novel threat. Each test used the same strongest AD, and a subset of the data was used to tune the loss threshold for an event to be labelled anomalous; this subset of data was not used in the final evaluation of the two-model detector's performance. Events labelled as anomalous were then passed to the appropriate novelty detector given the choice of held-out attack, and the overall precision, recall, and AUC were computed taking into account the events rejected by the AD (\ref{fig:e2e_results}). The results closely track the results from the novelty detector due to the near-perfect performance of the AD. Averaged evenly over all attack classes, two-model approach demonstrated $ 83\% $ precision and $ 63\% $ recall. However, it is worth noting here that almost none of the wrongly classified events were benign events due to the strong performance of our AD, but rather were anomalous events that were mistakenly identified as novel. Furthermore, \ref{fig:overall_results} shows the significant boost in performance our approach provides compared to traditional approaches discussed previously. 

\begin{wrapfigure}{r}{0.5\textwidth}
    \caption{Average results for the single autoencoder, dual autoencoder, and dual autoencoder with graph features, trained on MAWI and NCCDC and tested on NCCDC. Note: The proprietary Codex and DIF datasets are excluded.}
    \centering
    \begin{tabular}{|c|c|c|c|}
    \hline
         & AUC & Precision & Recall \\
\hline
Single & 0.69 & 0.63 & 0.64 \\
Dual & 0.83 & 0.84 & 0.76 \\
Dual with Graph & 0.93 & 0.96 & 0.93 \\
\hline
    \end{tabular}
    \label{fig:overall_results}
\end{wrapfigure}

\section{Feature Selection and Hyperparameter Tuning}

We found that a larger feature set containing 6 flow-level features and all graph features was required to obtain strong performance with the AD, whereas only a subset of those features were required to achieve similar performance with the Novelty Detector. Feature selection for each model was conducted using a grid search over all possible feature combinations and performance was judged by the AUC of a holdout attack type. Furthermore, we found that normalization was essential to allowing our models to generalize between different networks.

We also conducted several experiments to determine a high performing structure for each of our AE involving modifications to the depth of the Autoencoder, the width of its latent space layer, activation functions for the AE layers, dropout, and batch normalization layers. Ultimately, both our AEs utilized a symmetric structure where successive layer widths increased or decreased by a fraction of 1.4 and the latent space layer contained 6 perceptrons.  Our experiments also found that the addition of dropout or batch-normalization layers between fully-connected layers led to a significant drop in performance and the Rectified Linear Unit (ReLU) activation function yielded the strongest performance.

\section{Conclusion and Future Work}
The novel Dual AE Approach, augmented with asset-level graph features, to ZDT detection proposed in this paper offers a viable approach to identifying unknown threats effectively. Compared with single-AE based approaches or modeling approaches based on just network flow features, our models demonstrate a superior ability to generalize detection to unseen network topologies without retraining. This provides the distinct benefits of being extremely cost effective to deploy and drastically reducing time to value for security operators using our models. The strong performance of our models and low false positive rates offer operators reliable detections and reduced alert fatigue. Furthermore, the features used in our models, such as flow duration, source and destination features, have been specifically chosen for maximum compatibility with existing security monitoring infrastructure and should be easily available on client networks without the overhead of installing specialized sensors. This flexibility allows for greater versatility of our models. The AEs and pipeline have also been tested with large-scale datasets from real organizational networks in addition to attack-range datasets to determine that our model can produce reliable detections based on security telemetry from networks that have thousands of hosts. These factors combined make our architectures and methodology outperform similar approached to ZDT detection while maintaining feasibility to deploy and operate.  

The ZDT detection pipeline that has been proposed has strong performance in differentiating benign from anomalous data. However, the performance of the model drops as the number of attack types increases (as seen by comparing the results of figures 4 and 5, with and without the Codex data), so we will continue to improve the performance of the Novelty Detector independently in order to improve the effectiveness of detections overall. Possible improvements could include changes to the model architectures or training methodology. We will also continue to train and test our models on datasets and networks with a large diversity in topologies, behaviours, and most importantly, attack types. We will also focus on creating architectures for continuous improvement of our models based on real-time operator feedback to improve the value our models bring to real-world detection and response for Security Operations Centers. 

\bibliographystyle{IEEEtran}
\bibliography{ref}

\end{document}